\documentclass[journal]{IEEEtran}
\usepackage{subcaption} 
\usepackage{cite}
\usepackage{amsmath}
\usepackage{array}
\usepackage{fixltx2e} 
\usepackage{cleveref}

\begin{document} 
\title{Finite-Key Analysis for Multiple-Antenna 
Quantum Key Distribution with Continuous Variables}  
\author{Masoud~Ghalaii, Ahmed~Q.~Lawey, and~Mohsen~Razavi
\thanks{MG is with the School of Computing and Mathematics, Manchester Metropolitan University, Manchester M1 5GD, United Kingdom. 
AQL and MR are with the School of Electronic and Electrical Engineering, University of Leeds, Leeds LS2 9JT, United Kingdom. 
Part of this work is presented at 24th International Conference on Transparent Optical Networks (ICTON, 2024). This work has been partly funded by the EPSRC via the CHEDDAR Project (Grant No. EP/X040518/1 and No. EP/Y037421/1) and by Cisco Research via Silicon Valley Community Foundation. All data generated in this paper can be reproduced by the provided methodology and equations. 
\textit{Corresponding author:} M. Ghalaii, e-mail: m.ghalaii@mmu.ac.uk.}} 

\maketitle 

\begin{abstract}  \noindent 
Multiple-input multiple-output (MIMO) techniques, known for their use in wireless communications, can also find applications in quantum communications systems. This is, for instance, relevant to satellite-based quantum communications when atmospheric turbulence and the time-varying nature of the channel can make their implementation difficult. 
Here, we study the application of MIMO techniques to continuous-variable (CV) quantum key distribution (QKD) in the realistic setting where a finite number of signals are exchanged. 
We show how data collected during the brief overpass time of low-earth-orbit satellites over ground stations can be processed to estimate MIMO channel parameters, extract a secret key, and quantify the key generation rate. 
For $2\times 2$ MIMO settings with Gaussian encoding and heterodyne detection, our finite-key analysis indicates that MIMO CV-QKD could outperform both multiplexed (with a nominal 3~dB improvement) and single-mode (with a nominal 6~dB nominal improvement) scenarios. 
Furthermore, we conjecture that the ultimate enhancement of a multiple-antenna CV-QKD system can be characterized by the product of the number of its inputs and outputs. 
\end{abstract} 

\begin{IEEEkeywords}
Quantum communications, cryptography, MIMO communications, QKD, continuous variables. 
\end{IEEEkeywords}

\IEEEpeerreviewmaketitle

\section{Introduction} \noindent  
Quantum key distribution (QKD) was introduced in 1980s with the BB84 protocol~\cite{BB84,BB84:V2014}. 
This breakthrough established the foundation for quantum secure communications and was subsequently followed by the development of various quantum cryptographic protocols, including quantum conferencing and secret-sharing~\cite{Cleve:QSS1999,Ribeiro:QSS2010,Ottaviani:QConf2019,Grasselli:QSS2019,Pirandola:QConf2020}, 
quantum random number generators~\cite{Colbeck:QRNGs2012,Ma:QRNGs2016,Acin:QRNGs2016}, 
quantum digital signatures~\cite{Dunjko:QDS2014,Wallden:QDS2015,Collins:QDS2016,Roberts:QDS2017}, 
quantum bit commitment~\cite{Mayers:BitCom1997,Buhrman::BitCom2008,Kent:BitCom2012}, and quantum-safe  computing~\cite{Broadbent:SQC2009,Barz:SQC2012,Fitzsimons:SQC2017,Kashefi:SQC2017}. 
Such efforts are currently being pursued, shaping future quantum communications networks~\cite{Kimble:QInt2008,Azuma:QNet2016,Pirandola:QNet2016,Wehner:QNet2018,Azuma:QInt2019}.  
On a grand scale, QKD, as the most mature quantum technology today, serves as a quantum-based method enabling distant parties to securely exchange a secret key, even amidst potential interception~\cite{Pirandola:AQCrypt2020}. The two main classes of QKD are continuous-variable (CV) and discrete-variable (DV) techniques. 
The class of CV-QKD encodes information into quadratures of laser light~\cite{Weedbrook:RMP2012}. 
It relies on a distinct set of strategies including those based on squeezed states~\cite{Ralph:CVQKD-Sq1999,Hillery:CVQKD-Sq2000}, coherent states~\cite{Grosshans:PRL2002,Grosshans:Nat2003}, and thermal/noisy states~\cite{Usenko:CVQKD-Th2010,Weedbrook:CVQKD-Th2012,Ottaviani:CVQKD-THz2020}. 
Other CV-QKD protocols involve no-switching~\cite{Weedbrook:CVQKD2004}, discrete modulation~\cite{Leverrier:DMCVQKD2009}, and two-way~\cite{Pirandola:CV2Way2008}. In contrast, the more explored class of DV-QKD exploits quantum resources with finite degrees of freedom, e.g.,~polarization of single photons, and includes the BB84~\cite{BB84,BB84:V2014}, E91~\cite{Ekert:1991}, B92~\cite{Bennett:1992} and six-state~\cite{Bruss:SixSt1998} protocols. There are also hybrid protocols where both CV and DV components are employed~\cite{Djordjevic:hybridQKD20,Primaatmaja:hybridQKD,Mycroft:hybridQKD}. 

Nevertheless, direct, or point-to-point, QKD connections (either CV or DV) encounter constraints regarding the rate at which secret keys can be exchanged relative to the distance, dictated by limits dependent on channel transmissivity and system noise. A rigorous account limits repeaterless QKD protocols to delivering up to about $1.44T$ secure bits per channel use over a link with transmissivity factor $T$ (a function of the distance of the link), a limit known as the PLOB bound~\cite{Pirandola:PLOB2017}. 
To boost the overall QKD reach, one may employ quantum repeaters~\cite{Briegel:QRs1998,Pirandola:QNet2019a} or establish satellite-based quantum links~\cite{Liao:SatQKD1,Liao:SatQKD2,Ren:SatQKD2017}. 

Another approach to improve the overall QKD secret key rate is to employ multiplexing and mode-diversity methods to establish several simultaneous channels for key distribution~\cite{Bahrani:2015,Sahu:MIMO2023,Koudia:Mplex2024,Ghalaii:MIMOConf}. These channels should ideally operate independently, thus leading to a linear augmentation of the key rate with the increase in the channel count. However, achieving this ideal scenario may pose challenges in certain situations. For instance, in satellite-based QKD, if we employ spatial multiplexing, phenomena such as beam scattering and atmospheric scintillation can lead to interference between multiplexed channels~\cite{Kodheli:CSatComm}. Similar challenges may arise in QKD implementations utilizing multi-core fiber, where cross-talk noise from neighboring channels can significantly degrade the key rate per individual channel mode~\cite{Dynes:MultiCoreQKD}. These scenarios mirror the issues encountered in well-examined multiple-input multiple-output (MIMO) channels in classical wireless communications~\cite{Arapoglou:MIMO_Rev2011,Heo:MIMO_Rev2023}. 


To address these challenges in the quantum domain, recent research has focused on integrating CV-QKD with MIMO techniques, which holds significant promise as it enables the retrieval of crucial phase and amplitude information essential for the QKD system operation. Specifically, for a MIMO CV-QKD channel, Ref.~\cite{Sahu:MIMO2023} derives the key rate in the asymptotic limit, where an infinite number of signals are exchanged. This work also examines the optimality of CV collective attacks in the MIMO context and demonstrates that multiplexing gain can be achieved using multiple transmitters and receivers, even in the presence of crosstalk between channels.

In this study, we further explore the potential of MIMO techniques to enhance QKD systems operating in challenging channel environments. 
We provide a more robust analysis of the system by considering a broader set of MIMO parameters and exploring various scenarios that may arise in such channels. 
One practical constraint in satellite-based communications is the limited overpass time of the satellite over the ground station, which restricts the number of signals that can be exchanged.
Therefore, we focus on the finite-key performance of MIMO CV-QKD systems, where only a limited number of signals—and thus a finite number of data samples—can be realistically collected.
We examine how a portion of the data must be allocated for parameter estimation and investigate how, and at what rate, the remaining data can be processed to extract a secure key.
Furthermore, our findings identify operational regimes in which MIMO QKD systems offer advantages over single-input single-output (SISO) and multiplexed QKD implementations.
We also analyze the system’s dependence on key parameters and describe how these can be estimated from corresponding measurable quantities.  

The paper is structured as follows: 
Section~\ref{sec:sysdesc} provides an overview and extends the security analysis of the MIMO CV-QKD protocol in the asymptotic limit.
Section~\ref{sec:FSKey} presents our finite-size analysis, including data post-processing, parameter estimation, and key rate computation for the protocol. 
Section~\ref{sec:numericalresults} discusses our numerical results and Section~\ref{sec:conclusion} summarizes our findings and concludes the paper.  
  
\section{MIMO CV-QKD: System Description} 
\label{sec:sysdesc} 
  
In a $m\times n$ MIMO CV-QKD protocol, with $m$ transmitting and $n$ receiving antennas, Alice and Bob are allowed to use more than one quantum channel in parallel. 
As an example, in a satellite-to-ground QKD scenario, the transmitting satellite telescopes can each independently modulate some raw data and send it towards ground stations.  
Depending on the location of the ground stations with respect to the focal point of the beam, each receiver could receive part of the transmitted signal from each transmitter. The collected signal at each receiver is measured using a heterodyne receiver. Alice and Bob would then classically communicate over an authenticated channel to extract a secret key out of the exchanged data points.

\begin{figure}[t]
\centering
\includegraphics[width=7.51cm]{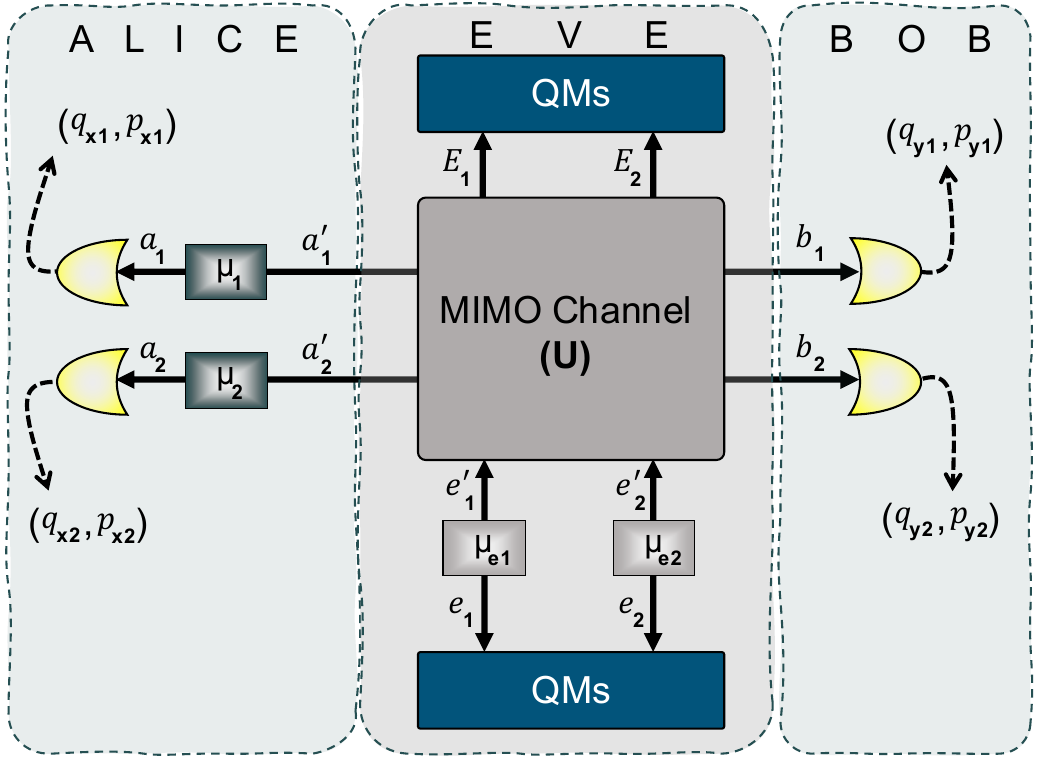}
\caption{\small Entanglement-based scheme of a $2\times 2$ MIMO CV-QKD protocols. Each $\mu$-box represents an entangled TMSV source and QMs are quantum memories. Heterodyne detection modules measure both quadratures of each optical mode at Alice and Bob labs.} 
\label{fig:MIMOsetup}
\end{figure}

For demonstration, the entanglement-based (EB) scheme for a $2\times 2$ MIMO CV-QKD protocol is schematically shown in Fig.~\ref{fig:MIMOsetup}.
Alice uses two zero-mean two-mode squeezed vacuum (TMSV) states, characterized by variances $\mu_1$ and $\mu_2$, to initiate the QKD task. Her entire state is then described by the following covariance matrix
\begin{align}
V_{A_1A_1'A_2A_2'} = \begin{pmatrix}
\mu_1 \mathbf{I} & k_1 \mathbf{Z} \\
k_1 \mathbf{Z} & \mu_1 \mathbf{I} \end{pmatrix}_{A_1A_1'} 
\oplus 
\begin{pmatrix}\mu_2 \mathbf{I}& k_2 \mathbf{Z} \\
k_2 \mathbf{Z}& \mu_2 \mathbf{I} 
\end{pmatrix}_{A_2A_2'}, 
\label{eq:CM_A}
\end{align}  
where $k_i=\sqrt{\mu_i^2-1}$, $\mathbf{I}=\text{diag}(1,1)$ and $\mathbf{Z}=\text{diag}(1,-1)$. $A_i$ and $A_i'$ describe optical modes that correspond to annihilation operators $\hat{a}_i=(\hat{q}_i+ i\hat{p}_i)/2$ and $\hat{a}_i'=(\hat{q}_i'+ i\hat{p}_i')/2$, respectively, where $\hat{q}_i$ and $\hat{p}_i$ are optical field quadratures. 
We have chosen the commutation relations $[\hat q_i,\hat p_i]=[\hat q_i',\hat p_i']=2i$ such that the variance of vacuum noise is 1 shot noise unit (SNU).

Alice sends modes $A_1'$ and $A_2'$ into the channel towards Bob, while retaining modes $A_1$ and $A_2$ for herself. 
By doing a heterodyne measurement on modes $A_1$ and $A_2$, modes $A_1'$ and $A_2'$ would be projected into coherent states, whose quadratures follow a Gaussian statistics. That is, from eavesdropper's (Eve's) point of view, Alice's outputs are Gaussian thermal states. 
Next, after channel's interaction that changes the covariance matrix $V_{A_1A_1'A_2A_2'}$ to $V_{A_1B_1A_2B_2}$, Bob measures modes $B_1$ and $B_2$. 
Each detection module is modeled by efficiency $\eta$ and electronic noise $\nu_{\rm el}$~\cite{Garcia-Patron:Imperfect_HOM}. 

For our security analysis, {\color{black} we assume a collective attack in which Eve applies a multi-mode extension of the entangling cloner attack~\cite{Garcia-Patron:OptAttack}. In this attack, Eve uses two TMSV states, characterized by variances $\mu_{e1}$ and $\mu_{e2}$, to interact with Alice's signals via a unitary transformation $\textsc{\bf U}$, while storing other modes in her quantum memories (QMs). This ensures having a Gaussian channel, which is known to allow for the optimal attack by Eve in the asymptotic case. For a $2\times 2$ MIMO channel, $\textsc{\bf U}$  maps input annihilation operators in Fig.~\ref{fig:MIMOsetup} to output ones by $(\hat b_1 ~ \hat b_2 ~ \hat E_1 ~ \hat E_2)^T= \textsc{\bf U} (\hat a'_1 ~ \hat a'_2 ~ \hat e'_1 ~ \hat e'_2)^T$, where \cite{Sahu:MIMO2023}
\begin{equation}
\begin{aligned}
\textsc{\bf U}= 
\left(
\begin{array}{cc|cc}
u_{11} & u_{12} & u_{13} & u_{14} \\
u_{21} & u_{22} & u_{23} & u_{24} \\ \hline
u_{31} & u_{32} & u_{33} & u_{34} \\
u_{41} & u_{42} & u_{43} & u_{44}
\end{array}
\right).
\end{aligned}
\label{eq:U_matrix}
\end{equation}
We refer to the upper left block in $\textsc{\bf U}$ as the $\textsc{\bf H}$ matrix, which effectively replicates a classical MIMO channel.}


Not surprisingly, and differently from the conventional SISO case, MIMO channels may interfere with each other, which could result in correlation between output modes and their corresponding measurements. 
For reference, Appendix \ref{sec:asympkey} summarizes previous work on MIMO CV-QKD in the asymptotic limit.
In the following, we present a finite-key analysis of such systems, where the number of exchanged signals is finite due to constraints such as the brief overpass time of satellites over ground stations. We use the $2\times 2$ MIMO channel as the base for explaining our analysis. We will consider other cases in Sec.~\ref{sec:numericalresults}, where we present our numerical results.

\section{Finite-Key Analysis for MIMO CV-QKD}  \noindent 
\label{sec:FSKey}  
In the finite-size key setting, Alice and Bob need to apply specific data processing steps to their collected data. This section outlines these essential processing steps under the assumption of a limited number of exchanged signals. 

\subsection{Data Slicing}  
\label{sec:DataSlicing}
{\color{black} In QKD, we often estimate or bound key system parameters by collecting data over a period of time. For a nominal time-invariant channel, the larger the size of the collected data, the tighter would be the corresponding bounds. This is because, by the law of large numbers, the ratios measured in a real experiment approach better the corresponding probabilities that we wish to bound. In a time varying channel, such as the one we expect to deal with in the satellite-to-ground scenario, this long-term averaging may be disadvantageous as periods of good channel conditions may be canceled out by those of adverse conditions. In this scenario, we may be better off to split the data into shorter segments during which the channel is not expected to change substantially and calculate the key rate per time slot. This could overall improve our secret key generation rate.

In our system, we assume that the overpass by the satellite starts at time $t_0$ and continues until time $t_{n_{\rm b}}$. We then split this time period to $n_{\rm b}$ segments with duration $\Delta t= t_s - t_{s-1}$, $s=1,\dots,n_{\rm b}$; see Fig.~\ref{fig:dataslicing}. This slicing strategy would also help with error reconciliation, so that a block with uncorrectable errors is discarded. As we will see later, the benefits from MIMO techniques can better be reaped if the channel changes slowly with $\Delta t$. This needs to be balanced out with the statistical fluctuations caused by the finite block size. 


For demonstration, let us assume a LEO satellite at 500~km altitude. The flight time of such a satellite within 1 rad from zenith, where we assume the quantum communications task is carried out, is estimated to be about 200 seconds. This is the time that we slice into $n_{\rm b}$ segments. Note that one could alternatively adapt an angular slicing, where, assuming a maximum zenith angle, the satellite trajectory is correspondingly divided into $n_{\rm b}$ segments. 
The total number of signals that could be transmitted and received depends on the pulse rate at the transmitter. In CV-QKD, this parameter is often chosen by considering the bandwidth of the receiver. The clock rate in such systems is then often on the order of tens of megahertz, but it can get to GHz regime in high rate systems. For a nominal pulse rate of 100~MHz, the total number of runs (transmitter usage) during the flight time would be $20 \times 10^9$. 
Assuming a channel coherence time of 100~ms, we can divide the flight time to $n_{\rm b}=2000$ segments, each corresponding to $10^7$ runs of the transmitter. Using this set of data, we will then do both parameter estimation and key extraction per block.}


\subsection{Notation: $2\times 2$ case}  
In the following, we present our notation for a $2\times 2$ MIMO CV-QKD setting. Suppose $N$ denotes the number of runs per $\Delta t$-long block. In the $2\times 2$ scheme, in each block, Alice randomly samples $[x_1]_j$ and $[x_2]_j$, $j=1,\dots,2N$, according to zero-mean normal distributions 
with variances $\sigma_i^2=\mu_i-1$, $i=1,2$, respectively. 
She then partitions $[x_1]_j$s into pairs $([q_{x_1}]_k,[p_{x_1}]_k)$ and $[x_2]_j$s into $([q_{x_2}]_k,[p_{x_2}]_k)$, for $k=1,\dots,N$, which are described by random variables $Q_{A_1}$, $P_{A_1}$, $Q_{A_2}$, and $P_{A_2}$, correspondingly. 
Using these, in total, she can prepare $N$ coherent states $|\alpha_{1k} \rangle$, where $\alpha_{1k}:=([q_{x_1}]_k+i[p_{x_1}]_k)/2$, and $N$ coherent states $|\alpha_{2k} \rangle$, where $\alpha_{2k}:=([q_{x_2}]_k+i[p_{x_2}]_k)/2$, to be sent from her transmitters. 
A total of $2N$ coherent signals are then exchanged within each time slot, which correspond to a block size of $N_{\rm b}= 4N$. 

\begin{figure}[t]
\centering
\includegraphics[width=7.51cm]{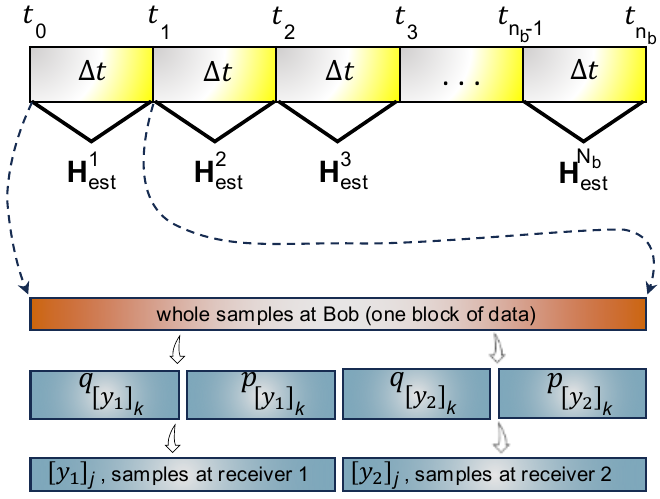} 
\caption{\small 
In a $2\times 2$ MIMO, Bob collects data within each data block (time interval) by measuring the four quadratures. The contribution of all quadrature values will form blocks of size $N_{\rm b}= 4N$, corresponding to $2N$ signals transmitted (received) at each input (output), of which $m$ will be consumed for parameter estimation and the computation of the secret key rate, and $n=N_{\rm b}-m$ will be used for key extraction. Alice's data will be sorted in the same manner and follow the same procedure (see main text).}
\label{fig:dataslicing}
\end{figure}

Corresponding to Alice's data points, as schematically shown in Fig.~\ref{fig:dataslicing}, Bob is also assumed to apply two heterodyne detection modules to obtain outcomes $([q_{y_1}]_k,[p_{y_1}]_k)$ and  $([q_{y_2}]_k,[p_{y_2}]_k)$, $k=1,\dots,N$. 
Similar to that of Alice, we define random variables $Q_{B_1}$, $P_{B_1}$, $Q_{B_2}$, and $P_{B_2}$ to represent random variables describing Bob's measurement outcomes. Similar to $[x_1]_j$ and $[x_2]_j$, the data form Bob's measurements can be paired together to define $[y_1]_j$ and $[y_2]_j$, $j=1,\dots,2N$.    

\subsection{Data Post-Processing}  
Data post-processing is an algorithmic procedure carried out by Alice and Bob on their respective data sets, with the goal of generating identical secret keys (binary strings) that remain unknown to Eve. The typical order of performing such a task is as follows: Parameter Estimation (PE), Error Correction (EC) or Reconciliation, and Privacy Amplification (PA)~\cite{Mountogiannakis:PRR2022a}. 
To account for the impact of the above procedures on the secret key generation rate in the finite-key regime, we start with the asymptotic key rate, $R_\infty$, where finite-key issues are not accounted for. By considering finite-size effects and the above post-processing steps, we then modify $R_\infty$ to derive a more precise and realistic estimate of the secret key generation rate. For instance, by performing the PE step, we obtain $R_\infty \rightarrow R_{\rm PE}$, where $R_{\rm PE}$ is determined by our bounds on the MIMO channel and the relevant noise parameters, consistent with the experimental observations and meeting the specified security requirements. In a similar fashion, by proceeding through the EC step, fed with $R_{\rm PE}$, we obtain $R_{\rm PE} \rightarrow R_{\rm EC}$, where $R_{\rm EC}$ depends on several error-correcting parameters, including those related to discretization, splitting, and encoding-decoding. 
Finally, by applying the privacy amplification step, we obtain 
$R_{\rm EC} \rightarrow R_{\rm PA} \equiv R$, which is our final secret key generation rate. 

By comparing the data post-processing steps of MIMO and SISO protocols, we note that PE may differ and involve more elaborate procedures in the MIMO case. This is due to the fact that several types of loss and noise parameters, including crosstalk ones, see \eqref{eq:U_matrix}, are required to be estimated in MIMO, whereas only one type of loss and one type of noise are involved in SISO. 
Nevertheless, one could argue that EC and PA could follow the same lines as in SISO. This is because Alice (Bob) could concatenate the data transmitted from (measured at) both transmitters (receivers), such that EC and PA for MIMO could be reduced to that of SISO. More precisely, while performing PE, we assume the concatenation of strings $Q_{A_1} P_{A_1}=X_1$ and $Q_{A_2} P_{A_2}=X_2$ for Alice and strings $Q_{B_1} P_{B_1}=Y_1$ and $Q_{B_2} P_{B_2}=Y_2$ for Bob. Next, we would assume the concatenation of the two strings of $X_1 X_2=X$ and $Y_1 Y_2=Y$ to respectively represent the entire data points retained by Alice and Bob. One could extend this method to more than two transmitters and/or receivers, though the PE step is expected to become more involved. 
Therefore, in the analysis we present in this study, we only elaborate on the PE stage for the $2\times 2$ MIMO setup in Fig.~\ref{fig:MIMOsetup} and refer to well-established techniques to carry out EC and PA, including Refs.~\cite{Mountogiannakis:PRR2022a} and \cite{Mountogiannakis:PRR2022b}. 

\subsection{Parameter Estimation for MIMO CV-QKD}   
\label{sec:PE_MIMO} 
{\color{black}  The first step in calculating the key rate is to account for the impact of parameter estimation. In the asymptotic case, summarized in Appendix \ref{sec:asympkey}, we can use a small percentage of the infinitely large dataset that Alice and Bob hold to accurately estimate the covariance matrix elements from which $R_\infty$ can be calculated. This can be done via the following expressions:
\begin{align}
R_{\infty}= \beta 
I_{\rm AB} - 
\chi_{\rm EB}, 
\label{eq:keyrate_inf}
\end{align}
where
\begin{align}
I_{\rm AB}= \frac{1}{2} \log_2\frac{\det{(\mathcal{A}+ \mathbf{I}_2)} \det{(\mathcal{B}+ \mathbf{I}_2)}}{\det{(V_{\rm AB}+\mathbf{I}_2\otimes \mathbf{I}_2)}} 
\end{align} 
is the mutual information between Alice and Bob, and the Holevo information term is given by
\begin{align}
\chi_{\rm EB}= \sum_{i} h((\zeta_i -1)/2) - \sum_{j} h((\zeta_{|j} -1)/2), 
\end{align}
where $h(x):=(x+1)\log_2(x+1)-x \log_2 x$,  
$\zeta_i$s are the symplectic eigenvalues of the covariance matrix 
\begin{align}
V_{\rm AB}= & 
\left(\begin{array}{cc}
\mathcal{A} & \mathcal{C} \\
\mathcal{C}^T  & \mathcal{B}
\end{array}\right),
\label{eq:VAB}
\end{align}
and $\zeta_{|j}$s are the symplectic eigenvalues of the conditional covariance matrix (on Bob's measurement on the received modes) $V_{{\rm AB}|\{ y_1,y_2\}} =\mathcal{A} - \mathcal{C} (\mathcal{B}+\mathbf{I}\otimes \mathbf{I})^{-1} \mathcal{C}^T$~\cite{Weedbrook:RMP2012}; see Appendix \ref{sec:asympkey} for further detail. This gives us a lower bound on the key rate based on the optimality of Gaussian attacks.

When we move to the finite-key regime, we can no longer accurately estimate the elements of $V_{\rm AB}$. In this case, we have to estimate such elements by using a finite number of data points, which means that we can only specify intervals within which the parameters of interest would lie with a certain level of confidence. In this case, the secret key generation rate is bounded by
\begin{align}
R_{\rm PE}= \min(R_\infty) 
\label{eq:keyrate_PE}
\end{align}
where the minimization is taken over all possible values of the estimated parameters within their confidence interval. For simplicity, we use the estimated values to calculate $I_{\rm AB}$ but find the worst-case scenario, denoted by $\widehat{\mathbf{P}}_{\rm WC}$, to maximize the Holevo information term $\chi_{\rm EB}$. 

In our case, we use $m/4$ randomly chosen 4-tuples $(q_{x_1}, p_{x_1}, q_{x_2}, p_{x_2})$ and their corresponding received values $(q_{y_1}, p_{y_1}, q_{y_2}, p_{y_2})$ at Bob to estimate the elements of the covariance matrix. The covariance term between any two zero-mean quadratures $Q_a$ and $Q_b$ is calculated as follows:
\begin{align}
\widehat{C}_{Q_aQ_b}:= \frac{1}{m/4} \sum_{k=1}^{m/4} [Q_a]_k [Q_a]_k, 
\label{eq:CQaQb}
\end{align}
where $Q_a,Q_b \in \{q_{x_1}, p_{x_1}, q_{x_2}, p_{x_2}, q_{y_1}, p_{y_1}, q_{y_2}, p_{y_2}\}$. 

In order to relate the confidence interval for an estimated parameter to the probability that the actual value of that parameter lies within that interval, we use the Gaussian attack assumption, which is known to be optimal in the asymptotic case. In effect, for any paired 4-tuples at Alice and Bob, we assume
\begin{align}
q_{y_1} = & \overline{u}_{11} q_{x_1} - \underline{u}_{11} p_{x_1} + \overline{u}_{12} q_{x_2} - \underline{u}_{12} p_{x_2} + \overline{z}_{y_1}, 
\label{eq:GaussChanModel_1} \\
p_{y_1} = & \underline{u}_{11} q_{x_1} + \overline{u}_{11} p_{x_1} + \underline{u}_{12} q_{x_2} + \overline{u}_{12} p_{x_2} + \underline{z}_{y_1}, 
\label{eq:GaussChanModel_2} \\
q_{y_2} = & \overline{u}_{21} q_{x_1} - \underline{u}_{21} p_{x_1} + \overline{u}_{22} q_{x_2} - \underline{u}_{22} p_{x_2} + \overline{z}_{y_2}, 
\label{eq:GaussChanModel_3} \\
p_{y_2} = & \underline{u}_{21} q_{x_1} + \overline{u}_{21} p_{x_1} + \underline{u}_{22} q_{x_2} + \overline{u}_{22} p_{x_2} + \underline{z}_{y_2}, 
\label{eq:GaussChanModel_4}
\end{align} 
where we denote $\overline{\circ}=\text{Re}\{\circ\}$ and $\underline{\circ}=\text{Im}\{\circ\}$ for convenience, and $\overline{z}_{y}$ and $\underline{z}_{y}$ represent the Gaussian noise variables that model Eve's contributions. Given that we are considering a collective attack, these noise variables are assumed to be independent from each other in different transmission rounds. Note that, in above equations, we assume $\eta=1$. In general, $\eta$ is a known constant and can later be used to appropriately scale the estimated values.

While $R_{\rm PE}$ can in principle be found by estimating the elements of $V_{\rm AB}$, it is common to define a set of equivalent physically relatable parameters that can tangibly model the worst-case channel compatible to our observed values. In the case of SISO, these are the total transmissivity of the channel and the excess noise parameter. In a similar fashion, in Ref.~\cite{Sahu:MIMO2023}, the authors have identified the following physically relatable parameters for the $2\times2$ MIMO scenario, which, in Appendix \ref{sec:asympkey}, we have shown their relationship to the covariance terms. These are the four components of the matrix $\textsc{\bf H}$, which define the MIMO channel transmissivity and crosstalk terms; $\xi_1$ and $\xi_2$, which are the observed excess noise values at Bob's first and second receiver, respectively; and $\xi_{\textsc{ct}}$ is a complex valued excess noise parameter that models correlation in the excess noise terms. 

In this work, we derive estimators for each of the above parameters and specify their confidence intervals with relation to the security parameters of the protocol. In particular, for each block in Fig.~\ref{fig:dataslicing}, we specify an estimate of the $\textsc{\bf H}$-matrix parameters denoted by
\begin{align}
\textsc{\bf H}_{\rm est}= 
\begin{pmatrix}
\widehat{u_{11}} & \widehat{u_{12}} \\
\widehat{u_{21}} & \widehat{u_{22}}
\end{pmatrix}, 
\label{eq:Hest_s}
\end{align}
and the relevant noise parameters. 
This is done by estimating the elements of the covariance matrix using \eqref{eq:CQaQb}, and finding the probability distribution of the estimator using the Gaussianity of \eqref{eq:GaussChanModel_1}--\eqref{eq:GaussChanModel_4}. In Appendix \ref{App:B}, we elaborate on the estimation of $u_{11}$ and $\xi_1$, as two representative examples. We show why they can be modeled as Gaussian random variables, and explain how to use $m$ data points to estimate their values, means, and variances. We further determine their worst-case scenario values by finding a confidence interval for each.}

\subsection{Finite-Key Rate for MIMO CV-QKD} 
\label{sec:FKrate}
Now that we have found $R_{\rm PE}$, we  account for the effect of error correction (reconciliation) and privacy amplification on the remaining part of the data, with size $n=N_{\rm b}-m$. Using the methodology and results reported in \cite{Mountogiannakis:PRR2022a, Mountogiannakis:PRR2022b}, we obtain the following key generation rate in the finite-size regime:
\begin{align}
 R 
  = & \frac{n p_{\rm EC}}{N_{\rm b}}
\Bigg(R_{\rm PE} - \frac{\Delta_{\rm aep}}{\sqrt{n}} + \frac{\Theta}{n} \Bigg),  
\label{eq:keynote_FS}  
\end{align}
where \cite{Pirandola:CompSec2021} 
\begin{align}
\Delta_{\rm aep}:= & 4\log_2(\sqrt{2^p}+2) \sqrt{\log_2(18 p_{\rm EC}^{-2} \varepsilon_{\rm s}^{-4} )} , \\
\Theta:= & \log_2\big[p_{\rm EC}(1- \varepsilon_{\rm s}^2/3)\big] + 2\log_2(\sqrt{2}\varepsilon_{\rm h}). 
\end{align}
In above, the total security parameter $\varepsilon$ is calculated by summing multiple error components: $\varepsilon= \varepsilon_{\rm sec} + \varepsilon_{\rm cor} + p_{\rm EC}(\varepsilon_{\rm ent} + 2\varepsilon_{\rm PE})$, where $\varepsilon_{\rm sec} =\varepsilon_{\rm h} + \varepsilon_{\rm s}$ defines the ``secrecy" of the PA state and includes a smoothing parameter $\varepsilon_{\rm s}$ and a hashing parameter $\varepsilon_{\rm h}$.
Also, assuming reverse reconciliation, Bob communicates a specific number of bits to Alice during hash comparison, given by $\log_2(1-\varepsilon_{\rm cor})$, which helps us bound the likelihood of disparity between Alice and Bob even if hash values match. 
This communication depends on the correctness parameter, $\varepsilon_{\rm cor}$, which quantifies the probability of potential differences. 
Finally, $\varepsilon_{\rm PE}$ is the overall error probability associated with PE and $\varepsilon_{\rm ent}$ is the error probability while bounding mutual information~\cite{Mountogiannakis:PRR2022a}. It is assumed that by using an analog-to-digital conversion, each continuous-variable symbol is encoded with $2^p$ bits of precision. 
In addition, An EC code may fail to correct errors within a block. This is captured by $p_{\rm EC}$, which is the probability that a block is successfully corrected. The frame error rate is then given by $\textsc{FER}=1-p_{\rm EC}$, which is the probability that the block is discarded. 


\section{Numerical Results} 
\label{sec:numericalresults}
\noindent
To highlight the potential of using MIMO techniques in QKD protocols, we carry out a numerical analysis in this section.
In particular, we focus on the relevant scenario of satellite-based CV-QKD, with multiple transmitting antennas on the satellite and multiple ground receivers.
First, we provide plots that facilitate a comparison between SISO and MIMO key generation rates in the asymptotic limit, revealing the ultimate performance of MIMO CV-QKD systems.
Next, we present our numerical results for the finite-key regime, assuming a limited number of exchanged signals.  

Here, we compute the secret key rate for specific MIMO channels depending on, e.g., how they affect signals and how close the transmitters/receivers are from each other. Assuming two line-of-sight (LOS) links between transmitters and receivers, one may adopt the diagonal terms $u_{11}=\sqrt{T_1}$ and $u_{22}= \sqrt{T_2}$, which state that signals leaving transmitter 1 (2) and reaching receiver 1 (2) have gone through a channel with transmissivity $T_1$ ($T_2$). 
On the other hand, the impact of signals leaving transmitter 1 (2) and reaching receiver 2 (1) is described by the crosstalk transmissivities  $\sqrt{T_{12}}$ ($\sqrt{T_{21}}$). 
In addition, in a coherent setting, one may consider the fact that the links may change the signals' phase from transmitter 1 (2) to receiver 2 (1), which are represented by $\theta_{12}$ ($\theta_{21}$).
The MIMO channel in this scenario can then be modeled by 
\begin{align}
\textsc{\bf H}= 
\begin{pmatrix}
\sqrt{T_1} & e^{i\theta_{12}} \sqrt{T_{12}} \\
 e^{i\theta_{21}} \sqrt{T_{21}} & \sqrt{T_2}
\end{pmatrix}. 
\label{eq:MIMOchan}
\end{align} 

For demonstration purposes, we further simplify our representative channel model to the following 
\begin{align}
\textsc{\bf H}= 
\begin{pmatrix}
\sqrt{T_\textsc{los}} & e^{i\theta_{12}} \sqrt{T_\textsc{ct}} \\
 e^{i\theta_{21}} \sqrt{T_\textsc{ct}} & \sqrt{T_\textsc{los}}
\end{pmatrix}, 
\label{eq:MIMOchanLOS}
\end{align} 
where, $T_\textsc{los}$ and $T_\textsc{ct}$ represent line-of-sight and crosstalk transmissivities, respectively. The channel in \eqref{eq:MIMOchanLOS} could be materialized in a downlink satellite scenario, where the receivers are relatively close to each other and far from the transmitters, so part of the signals could be collected by {\it unintended} receivers. 
{\color{black}  By assuming a fixed phase shift in the off-diagonal terms, we implicitly assume that the rate of change in the channel as compared to the block width $\Delta t$ is slow so that by averaging over $m$ random samples used in parameter estimation, we can recover the phase shift term. In the opposite regime, where channel (and especially the relevant phase shift) changes too quickly, the averaging performed in the PE stage is expected to remove any cross-correlation terms, in which case the off-diagonal terms of $\textsc{\bf H}$ would become zero. To capture this effect, and other cases in between,} we assume that $T_\textsc{ct}= \alpha T_\textsc{los}$, with $0\leq \alpha \leq 1$, noting that $\alpha=0$ reduces the MIMO channel to two independent multiplexed channels.

\begin{figure}
\begin{subfigure}[b]{0.53\textwidth} \centering
   \includegraphics[width=7.51cm]{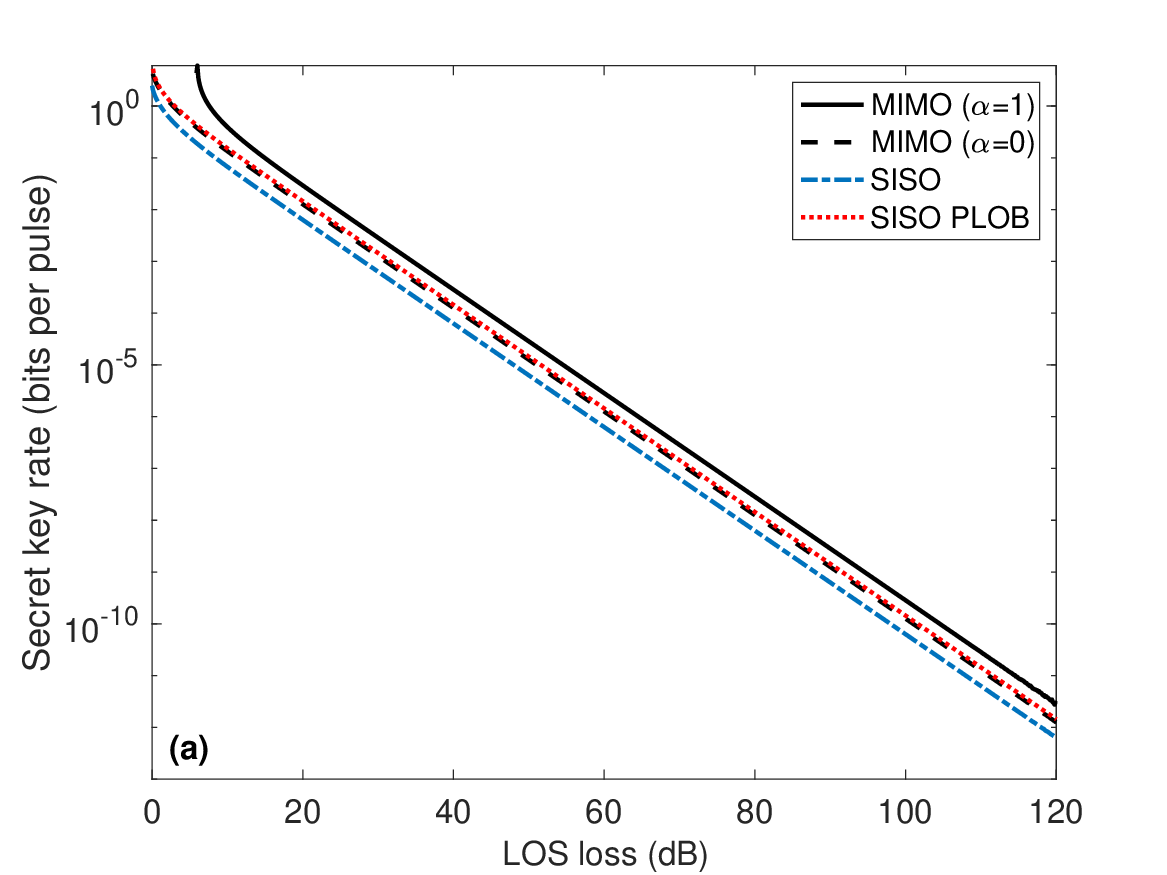} 
\end{subfigure}
\begin{subfigure}[b]{0.53\textwidth} \centering
   \includegraphics[width=7.51cm]{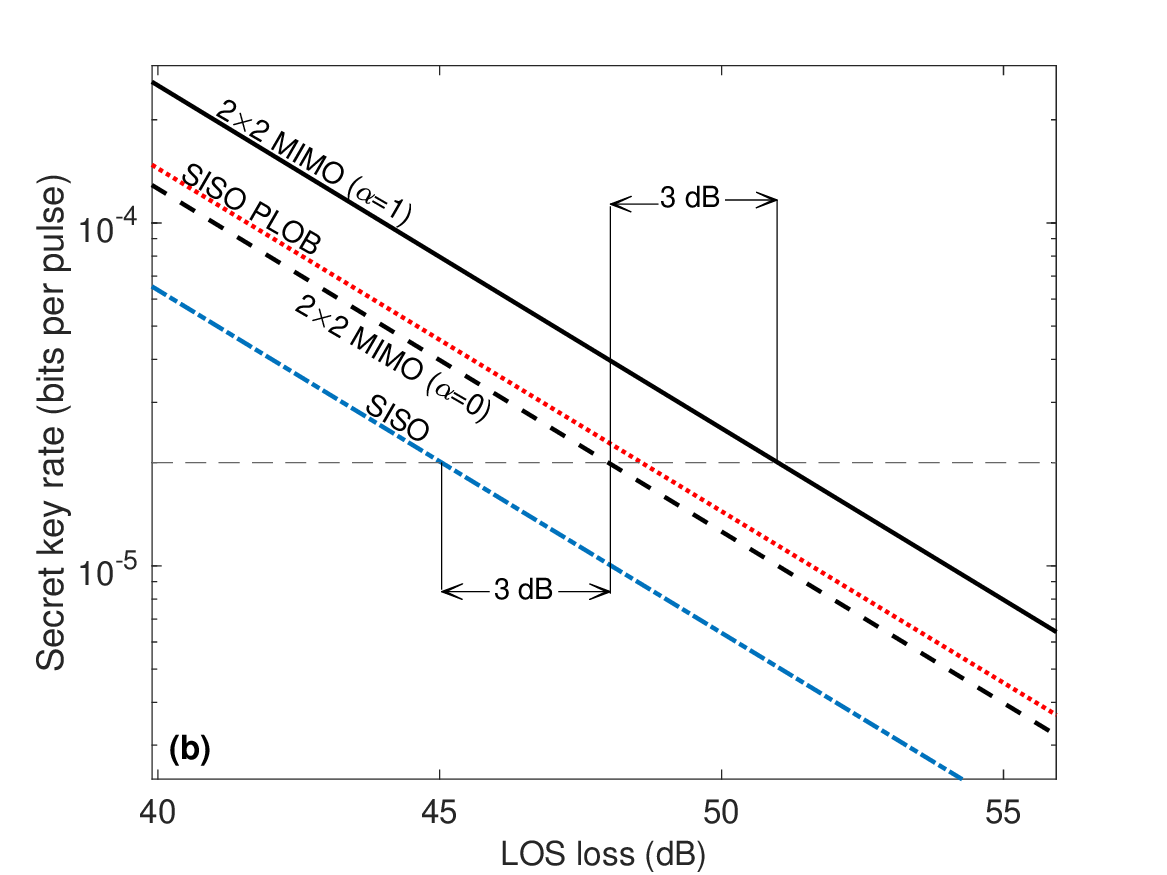}
\end{subfigure}
\caption{\small (a) The asymptotic secret key rate for $2\times 2$ MIMO CV-QKD versus line-of-sight loss, compared to the asymptotic rate of SISO CV-QKD, multiplexed rate (MIMO with $\alpha=0$), and the PLOB bound; (b) An area of (a) zoomed in. Here, we assume a pure-loss channel with no excess noise contribution from either the channel or detection modules {\color{black} ($\xi_1 = \xi_2 = \xi_{\textsc{ct}} = \nu_{\rm el} = 0$), receiver efficiency $\eta=1$, and reconciliation efficiency $\beta=1$.} In all cases, we have equal modulation variances $\mu_1=\mu_2=10$~SNU at both inputs. }
\label{fig:pureloss}
\end{figure}

Figure~\ref{fig:pureloss} shows the asymptotic secret key generation rate, $R_\infty$, achievable under different scenarios. The SISO PLOB bound \cite{Pirandola:PLOB2017}, given by $-\log_2(1-T_\textsc{los})$, gives the maximum key rate achievable for a SISO channel with transmissivity $T_\textsc{los}$. Here, we assume a pureloss MIMO channel with no noise contribution from either the channel or detection modules, $\theta_{12}=\theta_{21}=0$, as well as setting reconciliation efficiency $\beta = 1$. 
Though unrealistic, this gives the ultimate performance one would expect from a MIMO CV-QKD protocol. 

Our observations are as follows. First, it is noteworthy that the MIMO system at $\alpha = 1$ can outperform the SISO PLOB bound over a wide range of channel losses. 
This shows how MIMO systems can improve the overall key rate when the transmitted power from each transmitter has equally been collected by both receivers. This is something that we expect to happen when receivers are close to each other and the beam size at the receiver end is rather large, as expected, e.g., in the satellite link. 
Second, if this is not necessarily the case and the line of sight link carries more power than the side beam, i.e., $\alpha<1$, then the performance drops and at the limit of $\alpha = 0$ we get to the multiplexed scenario, where we can generate twice as many key bits as the SISO case when $\xi_\textsc{ct}=0$. 
Third, as shown in Fig.~\ref{fig:pureloss}(b), the multiplexed case exhibits a 3~dB advantage over the SISO rate---because of doubling the resources---while our $2\times 2$ MIMO system achieves a 3~dB gain over the multiplexed case.

\begin{figure}
\begin{subfigure}[h]{0.53\textwidth} \centering
   \includegraphics[width=7.51cm]{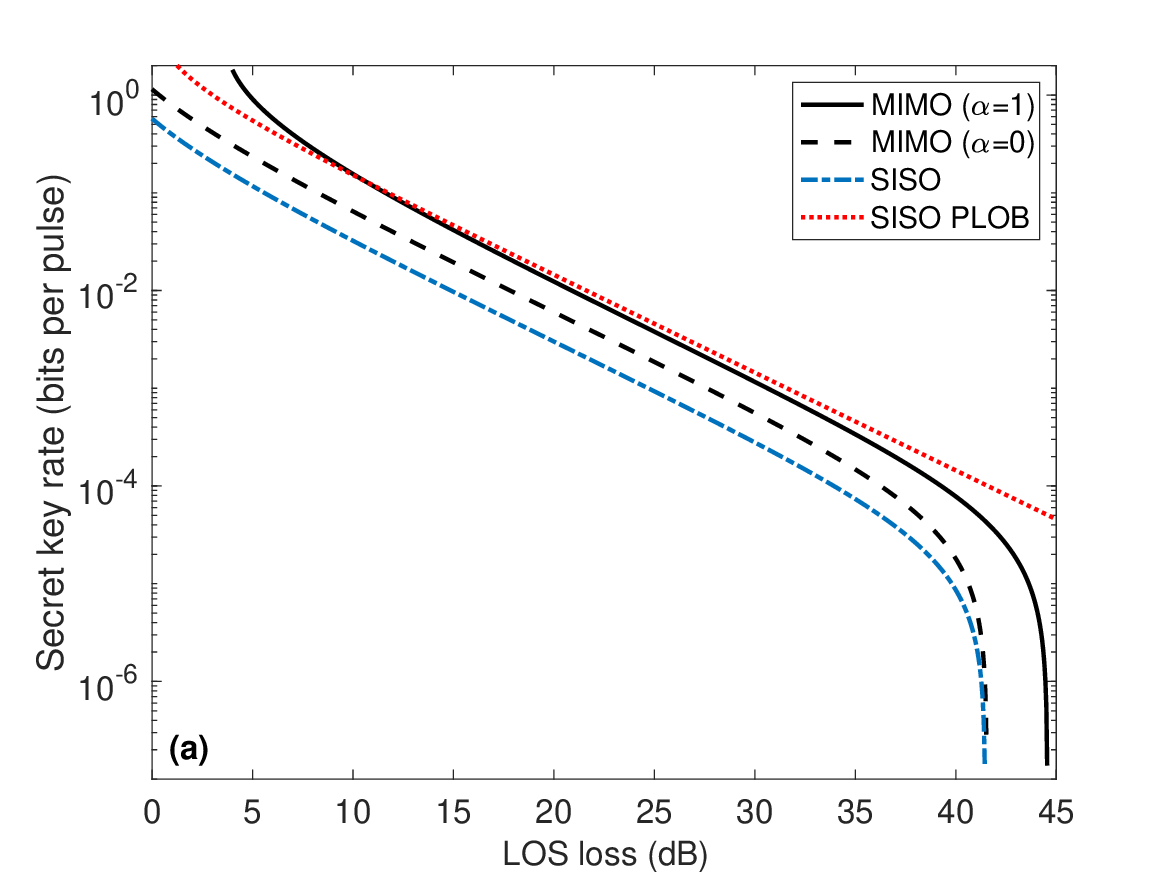}
\end{subfigure} 
\begin{subfigure}[h]{0.53\textwidth} \centering
   \includegraphics[width=7.51cm]{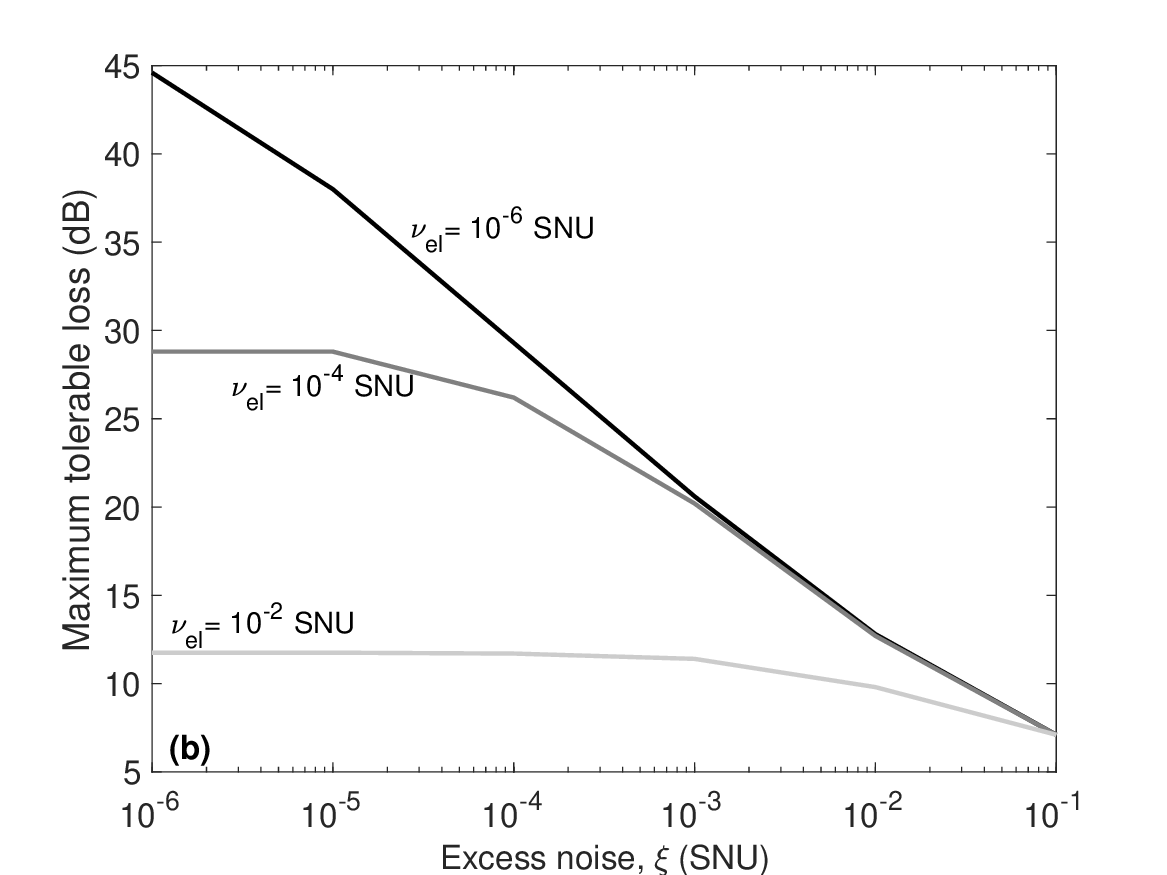}
\end{subfigure} 
\caption{\small Loss-noise trade-off in MIMO CV-QKD. (a) The asymptotic secret key rate for MIMO CV-QKD versus line-of-sight loss, compared to the asymptotic rate of SISO CV-QKD and the PLOB bound. Here, we assume a noisy channel with the receiver electronic noise ($\nu_{\rm el}$) and all types of channel excess noise ($\xi$'s) being equal to $10^{-6}$~SNU, receiver and reconciliation efficiencies are 
$\eta = 0.6$ and $\beta = 0.98$, respectively. 
(b) Maximum tolerable loss versus excess noise for a MIMO system with $\alpha=1$, considering different levels of electronic noise at the detectors. In all cases, equal modulation variances of $\mu_1=\mu_2=10$~SNU are applied at both inputs.}
\label{fig:lossynoisy}  
\end{figure}

\smallskip
{\em Channel Noise and Detection Noise.}  
Noise values, whether added by the channel, which is assumed untrusted, or during signal detection, which is assumed trusted, could also impact the performance. In Fig.~\ref{fig:lossynoisy}, for the same MIMO channel as given by \eqref{eq:MIMOchanLOS}, we present the curves for the same MIMO and SISO scenarios as in Fig.~\ref{fig:pureloss}. 
In our simulation, we have assumed the reconciliation efficiency $\beta = 0.98$, the detection efficiency $\eta = 0.6$, and the detection noise $\nu_{\rm el}= 10^{-6}$ SNU. 
In the case of SISO, the excess noise at the receiver is assumed to be $10^{-6}$ SNU, and to build the MIMO covariance matrix, we have assumed $\theta_{12}= \theta_{21}= \pi/2$, excess noise $\xi_1= \xi_2= 10^{-6}$ SNU, and crosstalk excess noise $\overline{\xi}_{\textsc{ct}}= \underline{\xi}_{\textsc{ct}}= 10^{-6}$ SNU. At each loss value, we have optimized the key rate of the MIMO protocol over Alice's modulation variance $\mu_1=\mu_2$.

We observe that, as shown in Fig.~\ref{fig:lossynoisy}(a), additional noise would restrict the reach of the QKD system. Nevertheless, the MIMO rate consistently outperforms the SISO rate across all $\alpha$ values. 
To further investigate the impact of electronic noise added at the receiver, Fig.~\ref{fig:lossynoisy}(b) plots the maximum tolerable loss against channel excess noise for several values of electronic noise. We conclude that the larger either type of noise, the lower the maximum tolerable loss. 
Additionally, more easily accessible values of electronic and excess noise---on the order of 0.001 SNU---can significantly reduce the tolerable loss to less than 20~dB, so it would be important to keep these noise terms as low as possible for satellite applications.

\smallskip 
{\em Finite-Key Rate.}  
We presented the theory of finite-key analysis for MIMO CV-QKD in Sec.~\ref{sec:FKrate}. We now extend our numerical results to the finite-key regime by assuming the specific MIMO channel given in \eqref{eq:MIMOchanLOS}. 
We study the effect of reducing the number of data sample points by decreasing the block size from infinity to a finite number. 
We observe that higher channel loss not only reduces the overall rate but also requires larger data block sizes to maintain a positive key rate. 

\begin{figure}
\begin{subfigure}[t]{0.53\textwidth} \centering
   \includegraphics[width=7.51cm]{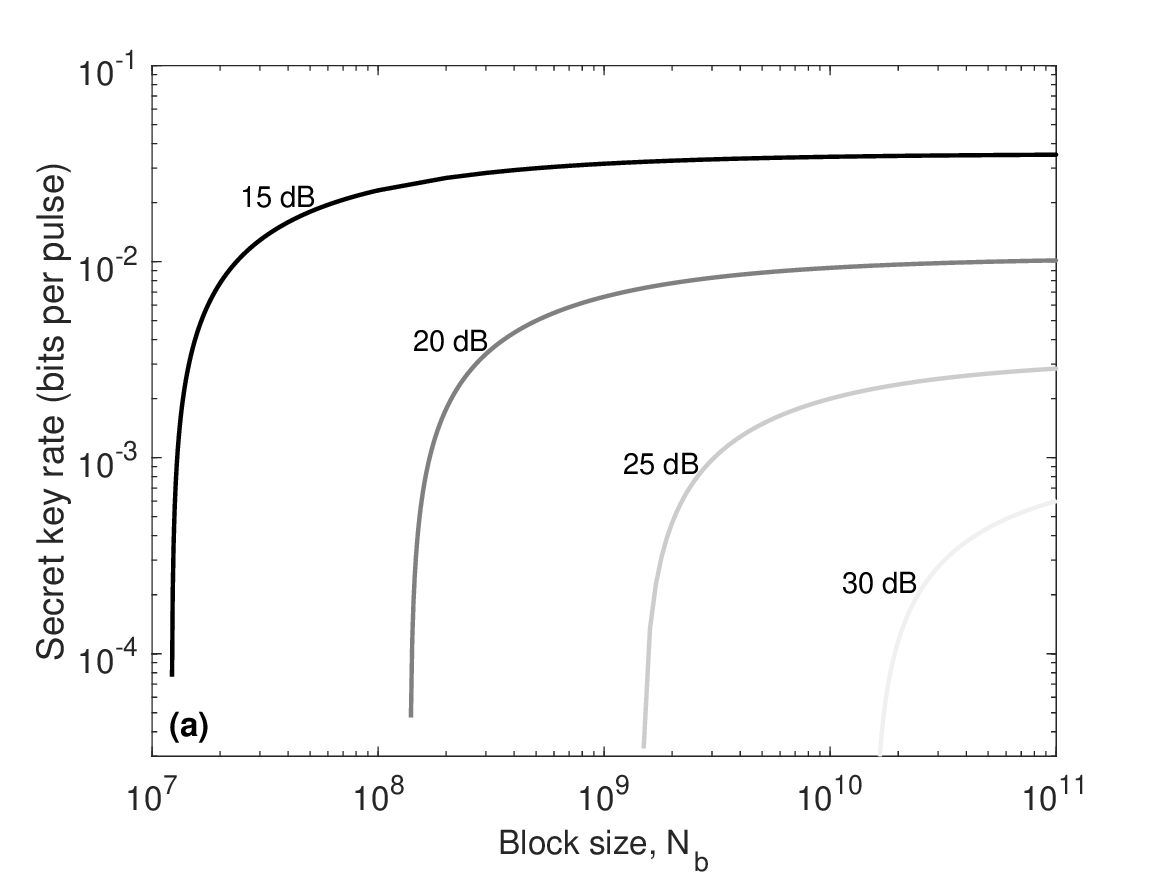} 
\end{subfigure}
\begin{subfigure}[t]{0.53\textwidth} \centering
   \includegraphics[width=7.51cm]{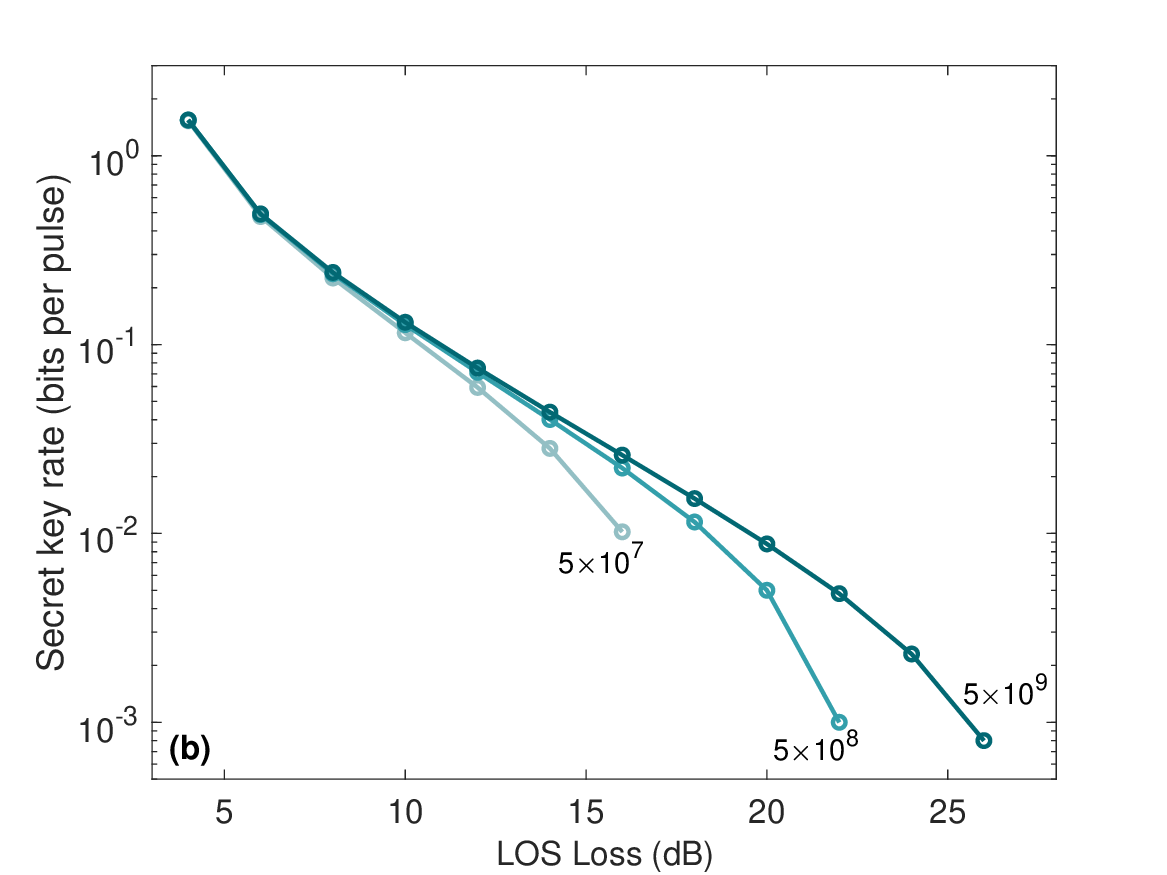}
\end{subfigure}
\caption{\small The secret finite-key rate for MIMO CV-QKD {\color{black}  with $\alpha=1$} versus (a) block size and (b) LOS loss. In (a), the numbers on each curve represent the LOS loss. In (b), they refer to the block size $N_{\rm b}$. Protocol parameters are: receiver electronic noise and all types of channel excess noise being equal to $10^{-6}$~SNU; receiver and reconciliation efficiencies are $\eta = 0.6$ and $\beta = 0.98$; Data processing parameters are $m=0.1 N_{\rm b}$, $p_{\rm EC}=0.95$, $p=5$; and post-processing errors are such that the overall security parameter $\varepsilon \simeq 10^{-10}$. 
} 
\label{fig:fs_rate}
\end{figure}

The finite-key rate is analytically given in \eqref{eq:keynote_FS} for a protocol with overall security parameter $\epsilon$, and is numerically plotted in Fig.~\ref{fig:fs_rate}.  
Figure~\ref{fig:fs_rate}(a) shows the finite-key rate for our $2\times 2$ MIMO CV-QKD protocol versus block size $N_{\rm b}$ for several values of LOS channel loss (15, 20, 25, and 30 dB). Figure~\ref{fig:fs_rate}(b) presents the finite-key rate versus LOS loss for two fixed block sizes. 
We assume that Alice and Bob consume 10\% of their collected data during parameter estimation ($m=0.1 N_{\rm b}$) and that their error correction successfully corrects errors with probability $p_{\rm EC}=0.95$. The protocol's overall security parameter is set to $\varepsilon = 10^{-10}$. Additional nominal parameters follow those used in Fig.~\ref{fig:lossynoisy}(a) and are detailed in the figure caption.

Several key observations emerge from these results. The MIMO system does not deliver any key at small block size regime, while the key rate it offers approaches the asymptotic rate at large enough block sizes.  
It is observed that a non-negative key rate can be achieved in high LOS loss regimes, up to 30 dB, provided a sufficiently large block size is employed. It is important to note that the required block sizes remain within the range of current data processing capabilities~\cite{Mountogiannakis:PRR2022a,Mountogiannakis:PRR2022b}. For instance, positive key rates for 15 dB and 20 dB LOS loss require minimum block sizes of approximately $2\times 10^7$ and $10^9$, respectively. These values, while substantial, are manageable with modern computational resources. {\color{black}  The more challenging problem would perhaps be to collect a sufficient number of data points during the coherence time of the channel. Following the discussion in Sec.~\ref{sec:DataSlicing}, we might need high-rate systems to ensure that this happens.}

Figure~\ref{fig:fs_rate}(b) illustrates the quantitative relationship between block size and loss tolerance. The system can accommodate an additional 10 dB of LOS loss by increasing the block size from $5\times 10^{7}$ to $5\times 10^{9}$, demonstrating a practical pathway for extending transmission distances through increased data collection. 
These finite-key results identify the requirements for the practical viability of MIMO CV-QKD systems. Below, we extend this discussion to setups with a larger number of antennas.

\color{black} 
{\em Conjecture on $m\times n$ MIMO Enhancement.}  
The above findings lead us to formulate a tentative conjecture regarding the achievable key rate advantage one might expect from an $m\times n$ MIMO CV-QKD architecture:

\smallskip
\noindent\textbf{Conjecture 1 ($m \times n$ Scaling).} \textit{An $m\times n$ MIMO CV-QKD system can offer $10 \log_{10}(m\times n)$ in dB advantage over a standard baseline SISO system under certain conditions.}
\smallskip

To support the plausibility of this scaling law, we present an alternative algebraic approach that recovers the 6~dB advantage of $2\times 2$ MIMO over SISO discussed earlier, and extend the results to  $m\times n$ MIMO configurations.

To facilitate this discussion, we note that we can  use singular value decomposition technique to write $\textsc{\bf H}=\textsc{\bf SVD}$, where $\textsc{\bf V}$ is a diagonal matrix. We can then use the transformations $X\rightarrow \textsc{\bf D}X=X'$ and $Y\rightarrow \textsc{\bf S}^{-1}Y=Y'$ to concisely rewrite \eqref{eq:GaussChanModel_1}--\eqref{eq:GaussChanModel_4}, in the $2\times 2$ case, as follows
\begin{align} 
\begin{pmatrix}
Y_1'  \\
Y_2' 
\end{pmatrix} =
\begin{pmatrix}
\lambda_1 & 0 \\
0 & \lambda_2
\end{pmatrix} 
\begin{pmatrix}
X_1'  \\
X_2' 
\end{pmatrix} +
\begin{pmatrix}
e_1  \\
e_2 
\end{pmatrix}. 
\label{eq:SVD}
\end{align}
Here, $e_1$ and $e_2$ are jointly Gaussian but not necessarily independent. However, if we assume $\xi_{\textsc{ct} = 0}$, we can then think of the channel as two independent SISO CV-QKD link with transmissivities $\lambda_1$ and $\lambda_2$. The rate of such a system would then be given by 
\begin{align}
R_{\rm \infty}^{2\times 2}= R_{\infty}^{1\times 1}(\lambda_1)+R_{\infty}^{1\times 1}(\lambda_2),     
\label{eq:MIMOSISO}
\end{align}
where superscripts refer to the type of the setting, and the argument represents the transmissivity of that setting. Here, for simplicity, we have ignored the impact of the excess noise, which is assumed to be zero. 
For the MIMO channel in \eqref{eq:MIMOchanLOS}, we have $\lambda_1=\sqrt{T_\textsc{los}}(1+\sqrt{\alpha})$ and $\lambda_2=\sqrt{T_\textsc{los}}(1-\sqrt{\alpha})$, which, at $\alpha = 1$, would reduce \eqref{eq:MIMOSISO} to $R_{\infty}^{2\times 2}= R_{\infty}^{1\times 1}(2\sqrt{T_\textsc{los}}) = R_{\infty}^{1\times 1}(\sqrt{4 T_\textsc{los}})$. This recovers the 6~dB advantage that we observed earlier in Fig.~\ref{fig:pureloss}.

Along the same lines, we can observe a trend for $m\times n$ MIMO scenarios, where the input power is distributed \textit{equally} between the line-of-sight and the crosstalk beams. In effect, let's assume our observed matrix $\textsc{\bf H}$ is an $m \times n$ matrix of ones multiplied by $\sqrt{T_\textsc{los}}$. If that is the case and the channel is noiseless, the SVD of the MIMO channel results in one eigenvalue of $\sqrt{m n T_\textsc{los}}$ and zero otherwise, which results in the following MIMO rate:
\begin{align}
R_{\infty}^{m\times n}=  R_{\infty}^{1\times 1}(\sqrt{m n T_\textsc{los}}). 
\end{align} 
This proves the possibility of having a factor of $m \times n$ MIMO advantage over the SISO case.

\begin{figure}
\centering
\includegraphics[width=7.51cm]{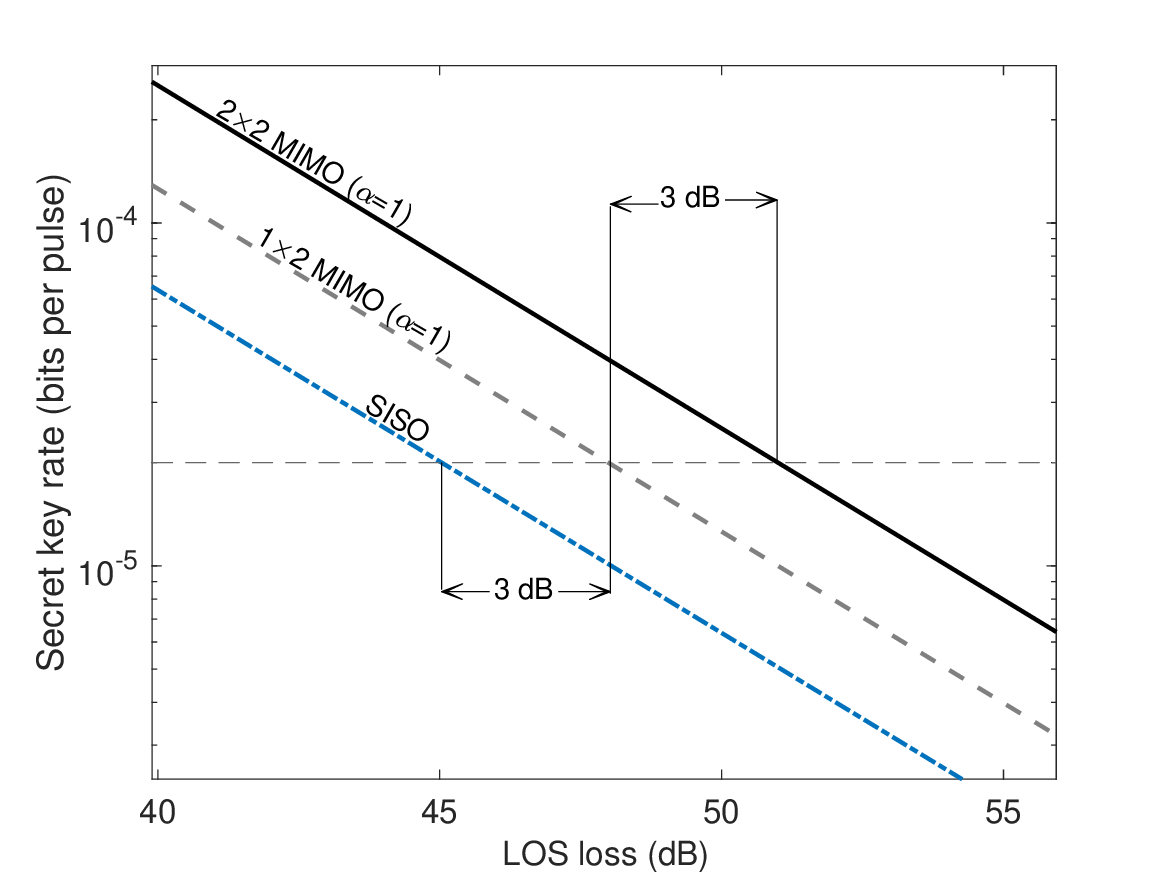}
\caption{The secret key rate for $1\times 2$ MIMO CV-QKD versus line-of-sight loss, compared to the asymptotic rate of SISO CV-QKD and the $2\times 2$ MIMO rate. Here, we assume $\alpha=1$ and a pureloss channel with no noise contribution from either the channel or detection modules, {\color{black}  receiver efficiency $\eta=1$ and reconciliation efficiency $\beta=1$}. In all cases, we have equal modulation variances $\mu_1=10$~SNU.} 
\label{fig:1by2MIMO}
\end{figure}

As an example, in Fig.~\ref{fig:1by2MIMO}, we compare the asymptotic performance of the $1\times 2$ MIMO system in a pure-loss scenario to that of the SISO and full $2\times 2$ MIMO configurations. We construct the $1\times 2$ MIMO CV-QKD system by choosing the vacuum state for Alice's second input in Fig.~\ref{fig:MIMOsetup}, i.e., by setting $\mu_2 = 1$. Consequently, the key rate formulation remains unchanged, allowing us to numerically compute the key rate. As we can see in Fig.~\ref{fig:1by2MIMO}, reducing the number of transmitters to one results in a 3~dB performance drop compared to the $2\times 2$ MIMO system, while still providing a 3~dB gain over the SISO baseline. One also notes that the key rate for $1\times 2$ MIMO matches the multiplexed rate, i.e., MIMO with $\alpha=0$. This means that, in a noise-free channel, the $1\times 2$ MIMO system with $\alpha=1$ could perform equally to two parallel links. 

It is important to note that the performance gain achieved above is for the special case where the observed $\textsc{\bf H}$ matrix has identical elements. This can potentially hold in slow-fading scenarios where receiver telescopes are close to each other, and the beams that reach the ground have widely diverged. This scenario may replicate nearly identical transmissivities but does not ensure identical phases for all terms. Nevertheless, a roughly similar gain advantage is expected in those scenarios as well. We also neglected excess noise in above calculations. Excess noise is not expected to change the scaling, although it would impact the maximum loss that can be tolerated. While all these claims need to be investigated more rigorously, the above results direct us to a promising approach in handling and implementing quantum communications systems in harsh channel conditions.



\color{black}

\section{Conclusions} 
\label{sec:conclusion}  
We explored the application and performance of multiple-input multiple-output techniques to continuous-variable quantum key distribution protocols. Our investigation used a $2 \times 2$ MIMO setting as a representative example to analyze the secret key rate in both asymptotic and finite-key regimes under Gaussian encoding and heterodyne detection. It was observed that we would most benefit from the MIMO structure once the power is equally distributed between the line-of-sight and the crosstalk beams. The results of our analysis also demonstrated a significant performance enhancement of MIMO CV-QKD compared to both multiplexed and single-mode CV-QKD scenarios. Specifically, our findings indicate a notable secret key rate improvement of 3 dB over multiplexed systems and a 6 dB improvement over single-mode systems.

Furthermore, our study provided insights into the ultimate enhancement of MIMO CV-QKD systems and characterized it by the product of the number of its inputs and outputs. This provides a valuable heuristic for scaling and designing future MIMO CV-QKD systems. 
For example, a MIMO setup with two transmitting and four receiving antennas can theoretically provide up to nearly 10 dB advantage over SISO systems---a substantial improvement when considering satellite-based QKD systems that typically experience 30-40 dB channel loss.  

We also conducted a comprehensive study on the influence of other critical parameters, including receivers' electronic noise and channel excess noise within the MIMO channels, on the overall performance. Our analysis revealed the robustness of the quantum MIMO technique against these imperfections, highlighting its practical viability. 
For most part, we considered a static scenario corresponding to the slow-fading regime. In practice, the channel parameters can change during the QKD signal transmission, whose impact can be countered by adaptive optics or better estimation techniques. The finite-key formulation provided in this work can then help us accurately estimate the achievable rate and best strategies for splitting the data.



\appendices

\section{An Overview of Asymptotic Analysis for MIMO CV-QKD}
\label{sec:asympkey} 
The asymptotic key rate of a CV-QKD protocol, where an infinite number of signals are transmitted and measured, in the reverse reconciliation case, is given by~\cite{Devetak-Winter}
\begin{align}
\label{eq:aym_R}
R_\infty= \beta I_{\rm AB} - \chi_{\rm EB},
\end{align}
where $\beta$ is the reconciliation efficiency, $I_{\rm AB}$ is the mutual information between Alice and Bob, and $\chi_{\rm EB}$ is the Holevo information between Eve and Bob. Quantities $I_{\rm AB}$ and $\chi_{\rm EB}$ can be  computed by using the covariance matrix of Alice's and Bob's modes, $V_{\rm AB}$, as we will introduce shortly. 
Assuming that Bob's (Alice's) data is described by variables $Y_1$ and $Y_2$ ($X_1$ and $X_2$), 
we have $I_{\rm AB}= I(X_1,X_2:Y_1,Y_2)$ and, for reverse reconciliation, $\chi_{\rm EB}= \chi({\bf E}:Y_1,Y_2)$, where ${\bf E}$ stands for all modes that belong to Eve.

\subsection{Covariance Matrix for $2 \times 2$ MIMO Channels} 
\label{sec:CM} 
Based on the transformation in \eqref{eq:U_matrix}, one can obtain the covariance matrix $V_{\rm AB}$ in \eqref{eq:VAB} where
\begin{align}
\mathcal{A}=& 
\left(\begin{array}{cc}
V_{A_1} & \mathcal{C}_A \\
\mathcal{C}_A^T & V_{A_2} 
\end{array}\right),  
\mathcal{B}= 
\left(\begin{array}{cc}
 V_{B_1} & \mathcal{C}_B \\
\mathcal{C}_B^T & V_{B_2} 
\end{array}\right),\quad \nonumber \\
&\mathcal{C}= 
\left(\begin{array}{cc}
\mathcal{C}_{A_1B_1} & \mathcal{C}_{A_1B_2} \\
\mathcal{C}_{A_2B_1} & \mathcal{C}_{A_2B_2} 
\end{array}\right). 
\label{eq:CM_MIMO}
\end{align}
In above, $V_{A_1}=\mu_1\mathbf{I}$, $V_{A_2}=\mu_2\mathbf{I}$, and $\mathcal{C}_A=0$; Matrix $\mathcal{C}$ contains correlation terms between Alice and Bob with 
$\mathcal{C}_{A_1B_1}=\sqrt{\eta} k_1 {\bf f}_{11}$,
$\mathcal{C}_{A_1B_2}=\sqrt{\eta} k_1 {\bf f}_{21}$,  
$\mathcal{C}_{A_2B_1}=\sqrt{\eta} k_2 {\bf f}_{12}$,
$\mathcal{C}_{A_2B_2}=\sqrt{\eta} k_2 {\bf f}_{22}$,
with 
\begin{align}
{\bf f}_{mn}=  \left(\begin{array}{cc}
\overline{u}_{mn} & \underline{u}_{mn} \\
\underline{u}_{mn} & -\overline{u}_{mn}
\end{array}\right), \quad\mbox{$m,n=1,2$}, 
\end{align}
where $\overline{\circ}=\text{Re}\{\circ\}$ and $\underline{\circ}=\text{Im}\{\circ\}$. The estimation of $\mathcal{C}$ would then enable us to estimate the $\textsc{\bf H}$-matrix parameters. 
Finally, the entries that define Bob's system are
$V_{B_1}=\delta_1 \mathbf{I}$, $V_{B_2}=\delta_2 \mathbf{I}$, and 
$\mathcal{C}_B=\begin{pmatrix}
\nu_1 & \nu_2 \\
-\nu_2 & \nu_1
\end{pmatrix}$, 
with
\begin{equation}
\label{eq:V_Bdelta1}
\delta_1=  \eta |u_{11}|^2 (\mu_1-1)+ \eta |u_{12}|^2 (\mu_2-1)+ \xi_1 + \nu_{\rm el}+ 1, 
\end{equation}
\begin{equation}
\label{eq:V_Bdelta2}
\delta_2=  \eta |u_{21}|^2 (\mu_1-1)+ \eta |u_{22}|^2 (\mu_2-1)+ \xi_2+ \nu_{\rm el}+ 1,  
\end{equation}
\begin{align}
\label{eq:V_Bnu1}
\nu_1 = & \xi_{\textsc{ct}1} + \eta \big(\overline{u}_{11}\overline{u}_{21} +\underline{u}_{11}\underline{u}_{21} \big) (\mu_1-1) \notag \\
& + \eta \big(\overline{u}_{12}\overline{u}_{22} +\underline{u}_{12}\underline{u}_{22} \big) (\mu_2-1), \\
\label{eq:V_Bnu2}
\nu_2 = & \xi_{\textsc{ct}2} + \eta \big(\overline{u}_{11}\underline{u}_{21} -\underline{u}_{11}\overline{u}_{21} \big) (\mu_1-1) \notag \\
& + \eta \big(\overline{u}_{12}\underline{u}_{22} -\underline{u}_{12}\overline{u}_{22} \big) (\mu_2-1). 
\end{align}
Having measured the $\textsc{\bf H}$-matrix parameters, the above covariance terms enable us to measure $\xi_1$ and $\xi_2$ as the observed excess noise values at Bob's first and second receiver, respectively. Also, $\xi_{\textsc{ct}1}$ and $\xi_{\textsc{ct}2}$ are the observed excess noise values added to Bob's cross-correlated data as captured by the $\mathcal{C}_B$ terms. One may also find it useful to make the complex number 
$\nu_1+i\nu_2= \eta u_{11}^*u_{21}(\mu_1-1)+ \eta u_{12}^*u_{22} (\mu_2-1)+ \xi_{\textsc{ct}}$, where  $\xi_{\textsc{ct}}:=\xi_{\textsc{ct}1} + i \xi_{\textsc{ct}2}$. 
This term is particularly interesting as there is no equivalent term in the SISO case. For a discussion on its impact the reader is referred to~\cite{Sahu:MIMO2023}. 




\section{Parameter Estimation: Finite-Size Key}
\label{App:B}
As we explained in Sec. \ref{sec:PE_MIMO}, here we outline the estimator for two representative parameters, $u_{11}$ and $\xi_1$, and find their respective probability distribution. Our methodology here is to express these parameters based on covariance matrix elements, where the latter can be estimated directly via \eqref{eq:CQaQb}.

\smallskip
{\em Estimation of MIMO Channel Parameters.}  Let us start by estimating $u_{11}$, whose real and imaginary parts appear in \eqref{eq:GaussChanModel_1} and \eqref{eq:GaussChanModel_2}, which we rewrite as follows 
\begin{align} 
q_{y_1} = & \overline{u}_{11} q_{x_1} + \overline{z}_{q_1}, 
\label{eq:GaussChanModel_1x}  \\
p_{y_1} = & \overline{u}_{11} p_{x_1} + \overline{z}_{p_1}, 
\label{eq:GaussChanModel_2x}
\end{align} 
where $\overline{z}_{q_1} := - \underline{u}_{11} p_{x_1} + \overline{u}_{12} q_{x_2} - \underline{u}_{12} p_{x_2} + \overline{z}_{y_1}$ and 
$\overline{z}_{p_1} := \underline{u}_{11} q_{x_1} + \underline{u}_{12} q_{x_2} + \overline{u}_{12} p_{x_2} + \underline{z}_{y_1}$ are zero-mean Gaussian variables. 
In the above equations, the real part of $u_{11}$ appears as the scaling factor that relates an Alice quadrature to that of Bob. By multiplying $q_{y_1}$ ($p_{y_1}$) by $q_{x_1}$ ($p_{x_1}$), and taking the average, we can estimate $\overline{u}_{11}$ in the following two ways:
\begin{align}
\widehat{\overline{u}_{11}} = \frac{\widehat{C}_{q_{x_1} q_{y_1}}}{\sigma_{q_{x_1}}^2} \quad \text{and} \quad 
\widehat{\overline{u}_{11}} = \frac{\widehat{C}_{p_{x_1} p_{y_1}}}{\sigma_{p_{x_1}}^2}, 
\end{align} 
where we have assumed that the covariance between Alice's variables as well as that between Alice's and Eve's variables are zero. 
To account for both sets of data equally, we define the following estimator for $\overline{u}_{11}$: 
\begin{align}
\widehat{\overline{u}_{11}} = & \frac{\widehat{C}_{q_{x_1} q_{y_1}}}{2 \sigma_{q_{x_1}}^2} + \frac{\widehat{C}_{p_{x_1} p_{y_1}}}{2 \sigma_{p_{x_1}}^2}. 
\end{align}
For simplicity, we further assume $\sigma_{q_{x_j}}^2=\sigma_{p_{x_j}}^2=\sigma_j^2$ and, hence, 
\begin{align}
\widehat{\overline{u}_{11}} = & \frac{\widehat{C}_{q_{x_1} q_{y_1}} + \widehat{C}_{p_{x_1} p_{y_1}}}{2 \sigma_1^2}. 
\label{eq:est_Ru11}
\end{align}

 
To find the statistics of $\widehat{\overline{u}_{11}}$, we should find those of $\widehat{C}_{q_{x_1} q_{y_1}}$ and $\widehat{C}_{p_{x_1} p_{y_1}}$, which due to the Gaussian channel assumption are normal variables with the following mean values
\begin{align}
E(\widehat{C}_{q_{x_1} q_{y_1}}) = \widehat{\overline{u}_{11}} \sigma_{q_{x_1}}^2, \quad
E(\widehat{C}_{p_{x_1} p_{y_1}}) = \widehat{\overline{u}_{11}} \sigma_{p_{x_1}}^2. 
\end{align} 
From \eqref{eq:est_Ru11}, and by assuming $\sigma_{q_{x_j}}^2=\sigma_{p_{x_j}}^2=\sigma_j^2$, we obtain the mean for $\widehat{\overline{u}_{11}}$ to be
\begin{align}
\overline{\mu}_{11}:= & E(\widehat{\overline{u}_{11}}) = \widehat{\overline{u}_{11}}. 
\label{eq:mubar_11}
\end{align}

Next, we find the variance of $\widehat{\overline{u}_{11}}$, denoted by $V(\widehat{\overline{u}_{11}})$. For this, we first develop the following noise estimators by using quadrature relations given in \eqref{eq:GaussChanModel_1x} and \eqref{eq:GaussChanModel_2x}:
\begin{align}
\widehat{\overline{\sigma}_{z_{q_1}}^2} = &  
\frac{1}{m/4} \sum_{k=1}^{m/4}
\Big( [q_{y_1}]_k - \overline{\mu}_{11} [q_{x_1}]_k \Big)^2, \\
\widehat{\overline{\sigma}_{z_{p_1}}^2} = & 
\frac{1}{m/4} \sum_{k=1}^{m/4}
\Big( [p_{y_1}]_k - \overline{\mu}_{11} [p_{x_1}]_k \Big)^2, 
\end{align}
where we used the estimated value for $\widehat{\overline{u}_{11}}=\overline{\mu}_{11}$. Such quantities appear in the computation of variances 
\begin{align}
V(\widehat{C}_{q_{x_1} q_{y_1}}) = & \frac{\overline{\mu}_{11}^2 \sigma_1^4}{m/4} \bigg(2+ \frac{\widehat{\overline{\sigma}_{z_{q_1}}^2}}{\overline{\mu}_{11}^2 \sigma_1^2} \bigg),  \\
V(\widehat{C}_{p_{x_1} p_{y_1}}) = & \frac{\overline{\mu}_{11}^2 \sigma_1^4}{m/4} \bigg(2+ \frac{\widehat{\overline{\sigma}_{z_{p_1}}^2}}{\overline{\mu}_{11}^2 \sigma_1^2} \bigg). 
\end{align} 
Now, similar to the mean, from \eqref{eq:est_Ru11}, we have  
\begin{align}
\overline{\sigma}^2_{11}:= & V(\widehat{\overline{u}_{11}}) = 
\frac{\overline{\mu}_{11}^2}{m} \bigg(4+ \frac{\widehat{\overline{\sigma}_{z_{q_1}}^2} + \widehat{\overline{\sigma}_{z_{p_1}}^2}}{\overline{\mu}_{11}^2 \sigma_1^2} \bigg).
\end{align} 

We therefore find that $\widehat{\overline{u}_{11}}$ has normal distribution with mean $\overline{\mu}_{11}$ and variance $\overline{\sigma}^2_{11}$. The confidence interval for $\widehat{\overline{u}_{11}}$, with an error probablity less than $\varepsilon_{\rm PE}$, is given by $\overline{u}_{11} - \overline{\Delta}_{11} < \widehat{\overline{u}_{11}} < \overline{u}_{11} + \overline{\Delta}_{11} $, where $\overline{\Delta}_{11} = \sqrt{2} \overline{\sigma}_{11} \cdot {\rm erf}^{-1}(1-\varepsilon_{\rm PE})$.  
The estimation of other components of the MIMO channel follows the same lines.  

\smallskip
{\em Estimation of Noise Parameters.}  
Our estimation of the key rate is based upon having the covariance matrix elements, which are a function of the MIMO channel components, i.e., $u_{ij}$s, as well as several noise parameters. 
In our analysis, such noise components appear at Bob's part of the covariance matrix and include the four parameters $\xi_1$, $\xi_2$, $\xi_{\textsc{ct}1}$ and $\xi_{\textsc{ct}2}$. In the following, we describe how such noise parameters could be estimated. 

As an example, let us focus on estimating $\xi_1$, which appears in the definition of $V_{B_1}= \delta_1 \mathbf{I}$, with $\delta_1= \eta  (\mu_1-1) |u_{11}|^2+ \eta  (\mu_2-1) |u_{12}|^2 + \xi_1$. 
Essentially, $\delta_1$ is the variance of both position and momentum quadratures $q_{y_1}$ and $p_{y_1}$. 
From \eqref{eq:GaussChanModel_1} and \eqref{eq:GaussChanModel_2}, one finds that the $\xi_1$ relates to both noise random variables $\overline{z}_{y_1}$ and $\underline{z}_{y_1}$, whose values could in general differ. 
However, 
we assume that their associated noise is the same and both resulting to an equal estimation of $\xi_1$. In any case, as in the estimation of $u_{11}$, one may estimate $\xi_1$ from both approaches and take an average value. Here, we focus on using $\overline{z}_{y_1}$. 

From \eqref{eq:GaussChanModel_1}, we have  
\begin{align}
\widehat{[\overline{z}_{y_1}]_k} =  [q_{y_1}]_k &- \widehat{\overline{u}_{11}} [q_{x_1}]_k + \widehat{\underline{u}_{11}} [p_{x_1}]_k 
- \widehat{\overline{u}_{12}} [q_{x_2}]_k\nonumber \\
&+ \widehat{\underline{u}_{12}} [p_{x_2}]_k, \quad\mbox{$k=1,\dots,m/4$},  
\end{align}
where $\widehat{{u}_{ij}}$s are estimated from the previous section. Then, we define the noise estimator
\begin{align}
\widehat{\xi_1} := \widehat{\sigma^2_{\overline{z}_{y_1}}} 
= & \frac{1}{m/4} \sum_{k=1}^{m/4} \Big( \widehat{[\overline{z}_{y_1}]_k} \Big)^2. 
\end{align}

To reveal some properties of our noise estimator, we see that the random variable $W_k:= [\overline{z}_{y_1}]_k/ \sigma_{\overline{z}_{y_1}}$, where $\sigma_{\overline{z}_{y_1}}$ is the standard deviation of $\overline{z}_{y_1}$, is $N(0,1)$ and that 
\begin{align}
\widehat{\xi_1}:= \widehat{\sigma^2_{\overline{z}_{y_1}}} 
= & \frac{1}{m/4} \sum_{k=1}^{m/4} W_k^2 
= \frac{\sigma^2_{\overline{z}_{y_1}}}{m/4} \chi_{m/4}^2(0), 
\end{align}
where $\chi_{m/4}^2(0)$ has a chi-square distribution, with $\kappa=m/4$ and $\lambda=0$.  
Thus, we conclude that the mean and the variance of our noise estimator are, respectively, given by 
\begin{align}
\label{eq:noise_1_est} 
E[\widehat{\xi_1}] := & \mu_{\xi_1}  =\frac{\sigma^2_{\overline{z}_{y_1}}}{m/4} \cdot (m/4) = \sigma^2_{\overline{z}_{y_1}}, \\
V[\widehat{\xi_1}] := & \sigma_{\xi_1}^2 = \bigg(\frac{\sigma^2_{\overline{z}_{y_1}}}{m/4} \bigg)^2 \cdot 2(m/4) = \frac{8 \sigma^4_{\overline{z}_{y_1}}}{m}. 
\end{align}
Moreover, since $\chi_m^2(0)$ approaches a normal distribution at $m \gg 1$, so does the distribution of $\widehat{\xi_1}$, such that we can use the confidence interval recipe we discussed for $u_{ij}$s. To conclude, $\widehat{\xi}_1$ is approximated as a Gaussian variable with the above mean and variance, and its estimation is accurate within an error probability less than $\varepsilon_{\rm PE}$, within the confidence interval $\xi_1 - \Delta_{\xi_1} < \widehat{\xi_1} < \xi_1 + \Delta_{\xi_1} $, where $\Delta_{\xi_1} = \sqrt{2} \sigma_{\xi_1} \cdot {\rm erf}^{-1}(1-\varepsilon_{\rm PE})$. 
 
\smallskip
{\em Worst-Case Scenario Parameters.} 
\noindent 
In the previous sections, we presented procedures for estimating the loss and noise parameters. We argued that each parameter can be modelled as a normally distributed variable, with its mean and variance estimated by sacrificing a portion of the data. As a result, each parameter lies within a confidence interval defined by a specified estimation error, $\varepsilon_{\rm PE}$. 

We now define a worst-case scenario parameter as the value that fall within its confidence interval and that yields the minimum key rate. Intuitively, we expect the worst-case scenario parameter for loss to be the one that gives the lowest transmissivities, such that, e.g., $[\widehat{\overline{u}_{11}}]_{\rm WC}= \overline{u}_{11} - \overline{\Delta}_{11} $, and the worst-case scenario parameter for noise to be the one that gives the largest amount, such that, e.g., $[\widehat{\xi_1}]_{\rm WC}=  \xi_1 + \Delta_{\xi_1} $. In any case, one can always optimize over the entire interval for each parameter to find a rigorous lower bound on the key rate. 

\ifCLASSOPTIONcaptionsoff
\fi 

\bibliographystyle{IEEEtran}
\bibliography{references} 

\end{document}